\documentclass[12pt]{article}
\usepackage{amsthm,amsmath,amssymb}
\usepackage{mathrsfs}
\usepackage{graphicx}
\usepackage{slashed}
\usepackage{multirow}
\usepackage{soul}
\usepackage[colorlinks, citecolor=blue,anchorcolor=blue,menucolor=blue,
linkcolor=blue,filecolor=blue,runcolor=blue,urlcolor=blue,frenchlinks=true]{hyperref}
\usepackage[normalem]{ulem}
\usepackage{xcolor}
\usepackage{outlines}
\usepackage[square, numbers, sort&compress]{natbib}
\usepackage[symbol]{footmisc}

\newcommand\pubdate{\today}

\def\Title#1{\begin{center} {\Large #1 } \end{center}}
\def\Author#1{\begin{center}{ \sc #1} \end{center}}
\def\Address#1{\begin{center}{ \small \it #1} \end{center}}

\newcommand\pubblock{\rightline{\begin{tabular}{l}  \\ 
         \pubdate  \end{tabular}}}
\newenvironment{Abstract}{\begin{quotation}  }{\end{quotation}}
\newenvironment{Presented}{\begin{quotation} \begin{center} 
             PRESENTED AT\end{center}\bigskip 
      \begin{center}\begin{large}}{\end{large}\end{center} \end{quotation}}

\begin{document}
\begin{titlepage}
 \pubblock
\vfill
\Title{Physical-Continuum Limit of the Nucleon Gluon Parton Distribution from Lattice QCD }
\vfill

\Author{William Good\footnote[1]{speaker}$^{1,2}$, Zhouyou Fan$^1$, Huey-Wen Lin$^{1,2}$}
\Address{$^1$Department of Physics and Astronomy, Michigan State University, East Lansing, MI 48824}
\Address{$^2$Department of Computational Math, Science, and Engineering, Michigan State University, East Lansing, MI 48824}

\vfill
\begin{Abstract}
    We present the key results from the first study of the $x$-dependent nucleon gluon distribution from lattice QCD extrapolated to the physical-continuum limit.  We use ensembles with $2+1+1$ flavors of highly improved staggered quarks (HISQ) generated by the MILC Collaboration with clover fermions for the valence action on three different lattice spacings and three different pion masses. We took up to $O(10^6)$ measurements of two-point correlators, obtaining good signal at boost momenta up to 3~GeV. We extrapolated the reduced pseudo-ITD matrix elements to the physical-continuum limit before extracting $xg(x)/\langle x \rangle_g$. Then using the gluon momentum fraction $\langle x \rangle_g$ calculated on the same configurations, we determine the unpolarized nucleon gluon PDF $xg(x)$ and compare results with other lattice and selected global fits.
\end{Abstract}

\vfill
\begin{Presented}
DIS2023: XXX International Workshop on Deep-Inelastic Scattering and
Related Subjects, \\
Michigan State University, USA, 27-31 March 2023 \\
     \includegraphics[width=9cm]{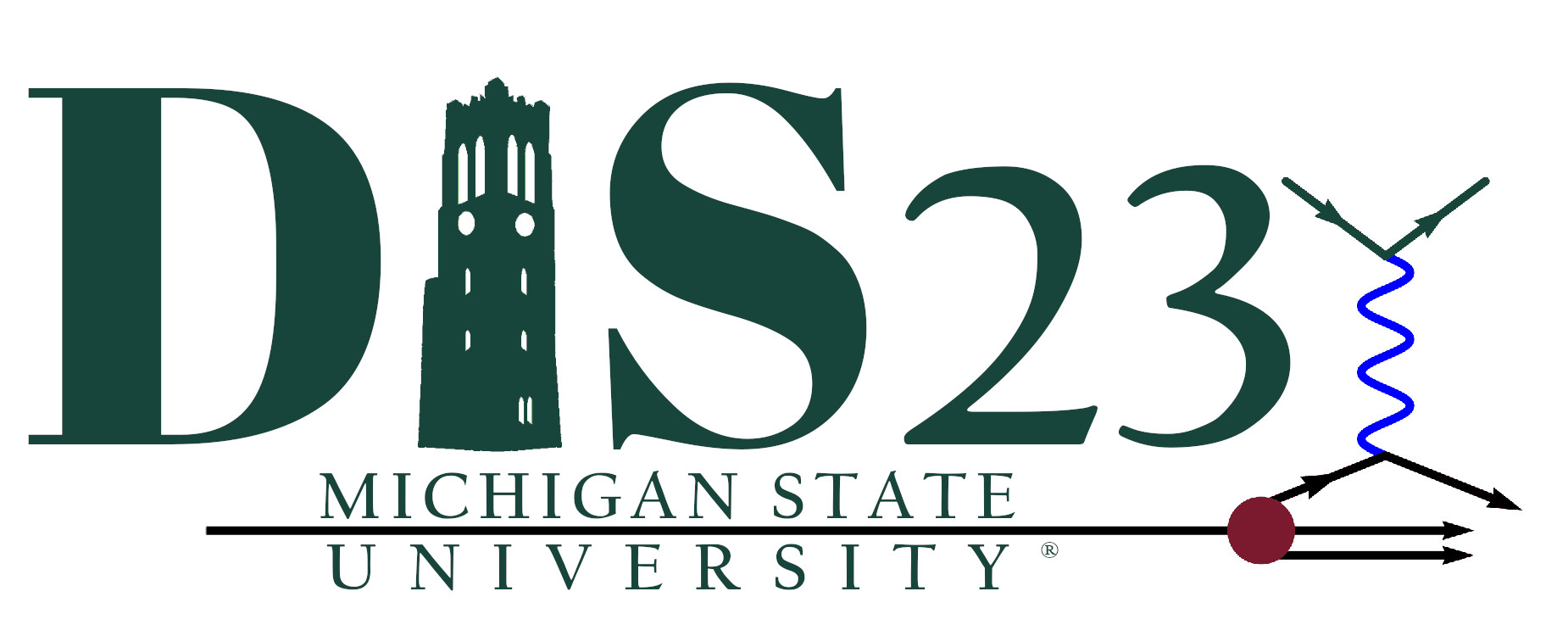}
\end{Presented}
\vfill
\end{titlepage}

\section{Introduction}

PDFs are nonperturbative function that quantify the probability of finding quarks and gluons in a hadron with some particular momentum fraction. The functions act as important input to many high-energy scattering experiments in hadron colliders~\cite{Harland-Lang:2014zoa,Dulat:2015mca,Abramowicz:2015mha,Accardi:2016qay,Alekhin:2017kpj,Ball:2017nwa,Hou:2019efy,Bailey:2019yze,Bailey:2020ooq,Ball:2021leu,ATLAS:2021vod}. In particular the gluon PDF $g(x)$ is required to calculate cross sections in $pp$ collisions including Higgs-boson production, jet production, and $J/\psi$ photoproduction~\cite{CMS:2012nga,Kogler:2018hem,mammeiproposal}. The future U.S.- and China- based electron ion colliders are planned to greatly increase our knowledge of the gluon PDF~\cite{Arrington:2021biu,Aguilar:2019teb,AbdulKhalek:2021gbh,Anderle:2021wcy}; however, in the meantime, our understanding is limited.

Lattice quantum chromodynamics (QCD) is a theoretical method for calculating nonperturbative QCD quantities that has full systematic control, which has been used since the advent of Large-Momentum Effective Theory to calculate $x$-dependent hadron structure~\cite{Ji:2013dva,Ji:2014gla,Ji:2017rah}. We specifically use the pseudo-PDF approach~\cite{Radyushkin:2017cyf} to calculate the unpolarized gluon PDF. Ref.~\cite{cichy2021progress} reviews recent developments in the use of lattice QCD for PDFs. In particular, it is very difficult to obtain good signal-to-noise ratios for gluon observables from lattice QCD and only a few exploratory studies have attempted this.

Here, we present highlights of the first continuum-limit unpolarized nucleon gluon PDF study. Namely, we briefly describe the lattice setup and extraction of the matrix elements from two point and three point correlators measured on the lattice. Then we describe the physical-continuum extrapolation of the reduced Ioffe-time pseudo-distribution before presenting the results of our light cone PDF matching compared to several other studies of the gluon PDF. The results presented here are given much further detail in our paper~\cite{fan2022gluon}.

\section{Lattice Setup, Correlators, and Matrix Elements}
We compute lattice calculations on four ensembles $N_f = 2+1+1$ highly improved staggered quarks (HISQ)~\cite{Follana:2006rc}, generated by the MILC Collaboration~\cite{Bazavov:2012xda}, with three different lattice spacings ($a\approx 0.9$, 0.12 and 0.15~fm) and three pion masses (220, 310, 690~MeV). We apply five steps of hypercubic smearing~\cite{Hasenfratz:2001hp} to the gauge links. Wilson-clover fermions are used in the valence sector, and the valence-quark masses are tuned to reproduce the lightest light and strange sea pseudoscalar meson masses (which correspond to light and heavy pion masses 310 and 690~MeV, respectively).
\begin{figure}[t!]
\centering
\includegraphics[width=0.41625\textwidth]{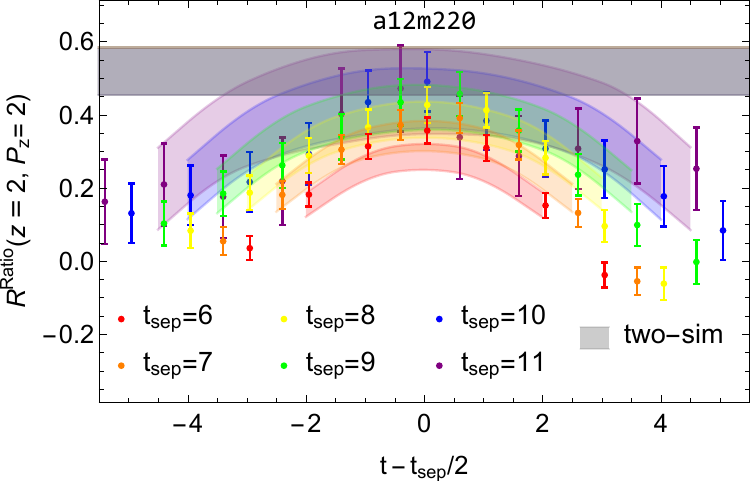}
\centering
\includegraphics[width=0.2475\textwidth]{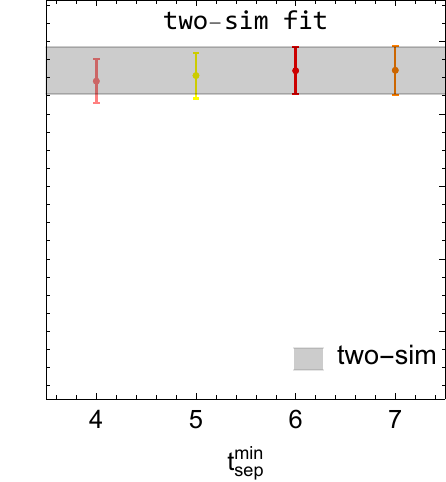}
\centering
\includegraphics[width=0.25\textwidth]{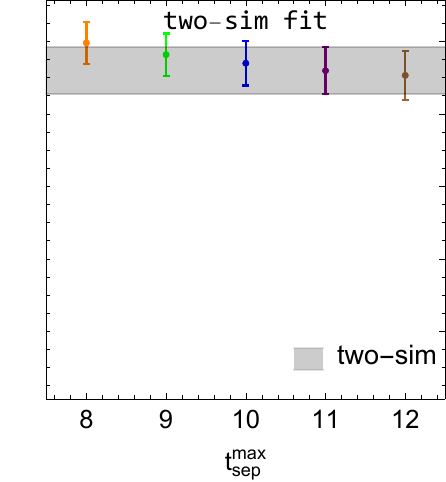}
\caption{
Example ratio plot (left) and two-sim fits (middle and right) of the light nucleon correlators at pion masses $M_\pi \approx  220$~MeV from the a12m220 ensemble.
The gray band on each plot gives the range for the ground-state matrix element extracted from the two-sim fits.
From left to right, the plots are:
the ratio of the three-point to two-point correlators and reconstructed two-sim fit bands using the final $t_\text{sep}$ inputs, shown as functions of $t-t_\text{sep}/2$,
the two-sim fit results using $t_\text{sep}\in[t_\text{sep}^\text{min},t_\text{sep}^\text{max}]$ varying $t_\text{sep}^\text{min}$, and the same procedure varying $t_\text{sep}^\text{max}$ instead. }
\label{fig:LRatio-fitcomp}
\end{figure}

The first step in obtaining the gluon matrix elements is to measure the two-point(2pt) and three-point(3pt) correlators on the lattice defined as follows:

\begin{equation}\label{}
C_N^\text{2pt}(P_z;t_\text{sep}) =
 \langle 0|\Gamma\int d^3y\, e^{-iyP_z}\chi(\vec y,t_\text{sep})\chi(\vec 0,0)|0\rangle,
\end{equation}

and

\begin{equation}\label{eq:3ptC}
 C_N^\text{3pt}(z,P_z;t_\text{sep},t) = 
 \langle 0|\Gamma\int d^3y\, e^{-iyP_z}\chi(\vec y,t_\text{sep}){\cal O}_g(z,t)\chi(\vec 0,0)|0\rangle,
\end{equation}

Here, $\chi$ is the nucleon interpolation operator, $\epsilon^{lmn}[{u(y)^l}^Ti\gamma_4\gamma_2\gamma_5 d^m(y)]u^n(y)$ (where $\{l,m,n\}$ are color indices, $u(y)$ and $d(y)$ are quark fields). $\Gamma=\frac{1}{2}(1+\gamma_4)$ is the projection operator. $P_z$ is the nucleon boost momentum in the $z$-direction. $t_\text{sep}$ is the source-sink separation time, and $t$ is the gluon-operator insertion time. ${\cal O}_g(z,t)$ is the gluon operator from Ref.~\cite{Balitsky:2019krf}:
\begin{equation}\label{eq:gluon_operator}
 {\cal O}_g(z)\equiv\sum_{i\neq z,t}{\cal O}(F^{ti},F^{ti};z)-\frac{1}{4}\sum_{i,j\neq z,t}{\cal O}(F^{ij},F^{ij};z),
\end{equation}
where the operator ${\cal O}(F^{\mu\nu}, F^{\alpha\beta};z) = F^\mu_\nu(z)U(z,0)F^{\alpha}_{\beta}(0)$, and $z$ is the Wilson link length. We used Gaussian momentum smearing~\cite{Bali:2016lva} on the quark field and took $\mathcal{O}(10^{5-6})$ measurements over $\sim$1000 lattice configurations for each ensemble to get improved signal out to a nucleon boost momenta of 3.0~GeV.

To obtain the ground-state matrix element, we used a two-state fit on the 2pt correlators, and a two-sim fit on the 3pt correlators: 

\begin{equation}
C_N^\text{2pt}(P_z,t_{sep}) = 
|A_{N,0}|^2 e^{-E_{N,0}t_{sep}} + |A_{N,1}|^2 e^{-E_{N,1}t_{sep}} + \ldots,
\label{eq:2pt_fit_formula}
\end{equation}
\begin{multline}
C_N^\text{3pt}(z,P_z,t,t_\text{sep}) = \\
 |A_{N,0}|^2\langle 0|O_g|0\rangle e^{-E_{N,0}t_\text{sep}} 
+|A_{N,0}||A_{N,1}|\langle 0|O_g|1\rangle e^{-E_{N,1}(t_\text{sep}-t)}e^{-E_{N,0}t} \\
+|A_{N,0}||A_{N,1}|\langle 1|O_g|0\rangle e^{-E_{N,0}(t_\text{sep}-t)}e^{-E_{N,1}t} 
+ |A_{N,1}|^2\langle 1|O_g|1\rangle e^{-E_{N,1}t_\text{sep}}
+ \ldots,
\label{eq:3pt_fit_formula}
\end{multline}
where the $|A_{N,i}|^2$ and $E_{N,i}$ are the ground-state ($i=0$) and first excited state ($i=1$) amplitude and energy, respectively.

We plot the ratio of the three-point to the two-point correlators
\begin{equation}\label{eq:ratio}
R_N(z,P_z,t_\text{sep},t)=\frac{C_N^\text{3pt}(z,P_z, t, t_\text{sep})}{C_N^\text{2pt}(P_z,t_{sep})}.
\end{equation}
Ideally, as $t_{sep}\rightarrow \infty$, $R_N \rightarrow \langle 0|O_g | 0 \rangle$ as defined in Eq.~\ref{eq:3pt_fit_formula}. Figure~\ref{fig:LRatio-fitcomp} shows one example of the ratio plots on the left, along with the fitted matrix element $\langle 0|O_g|0\rangle$ in the gray band. As expected, we see that as $t_sep$ increases, the ratio fits appear to monotonically approach our fitted matrix element. Additionally, the right two plots show that varying the range of $t_sep$ used in the fitting procedure does not change the matrix element significantly and that our choice of $t_{sep}$ is on a converging path. We see similar features on other samples of ratio plots in the paper \cite{fan2022gluon}.

\section{Results and Discussion}
\subsection{RpITDs and Continuum Extrapolation}
With our fitted ground-state matrix elements, we can compute the so called reduced Ioffe-time pseudo-distribution (RpITD)~\cite{Radyushkin:2017cyf,Orginos:2017kos,Zhang:2018diq,Li:2018tpe} 
\begin{equation}
\mathscr{M}(\nu,z^2)=\frac{\mathcal{M}(zP_z,z^2)/\mathcal{M}(0\cdot P_z,0)}{\mathcal{M}(z\cdot 0,z^2)/\mathcal{M}(0\cdot 0,0)},
\label{eq:RITD}
\end{equation}
where Ioffe time $\nu=zP_z$, and $\mathcal{M}(\nu,z^2)$ are the nucleon matrix elements at boost momentum $P_z$ and gluon operators with Wilson displacement $z$. By construction, the RpITD cancels renormalization of ${\cal O}(z)$ and kinematic factors. The ultraviolet divergences are removed, and the lattice systematics are reduced. RpITDs act as input into the pseudo-PDF framework to obtain PDFs \cite{Radyushkin:2017cyf}.

\begin{figure}[t!]
\centering
\centering
\includegraphics[width=0.46\textwidth]{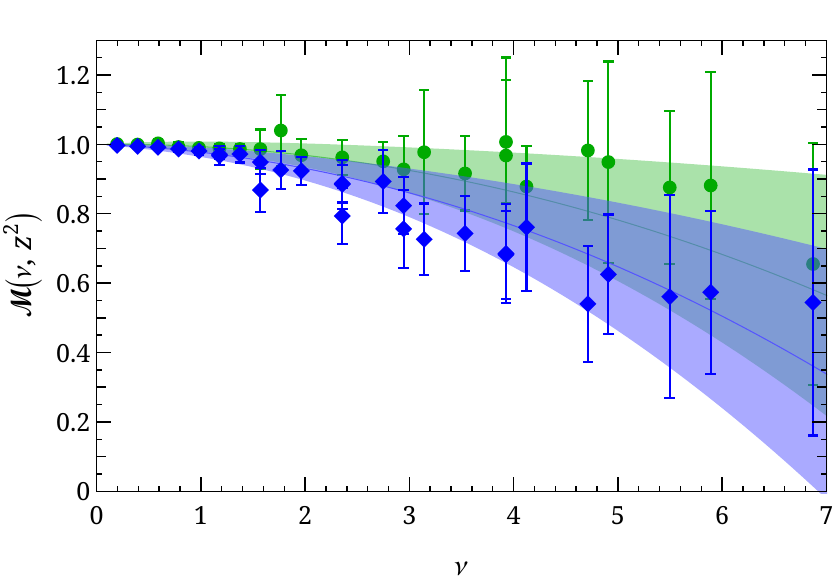}
\includegraphics[width=0.46\textwidth]{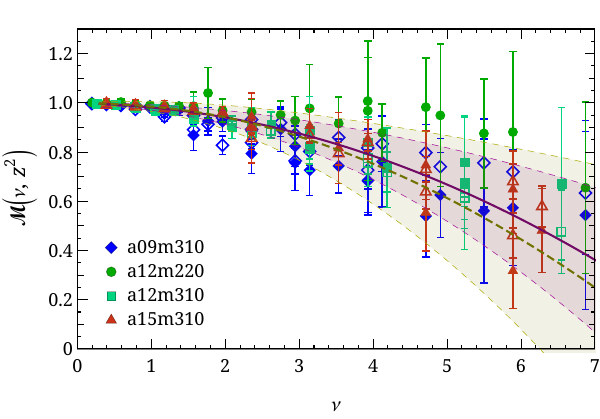}
\caption{
(Left) Examples of the RpITDs $\mathscr{M}$ reconstructed bands from fits in Eq.~\ref{eq:RpITD-extra} for a09m310 (blue points and light blue band), a12m220 (green) lattice ensembles. (Right) All data compared to the continuum extrapolation at the physical pion mass for $O(a)$ (dashed band) and $O(a^2)$ (solid band).
Open symbols indicate the data for the heavier quark mass of the ensembles with the same closed symbol.
}
\label{fig:RpITD_exta_1}
\end{figure}

We compute an extrapolation to physical pion mass and the continuum limit on the RpITD using the data from our four ensembles at the light and strange masses. We implement the following extrapolation form: 
\begin{align}\label{eq:RpITD-extra}
\mathscr{M}(\nu, z^2,a,M_{\pi})={}&\left(\sum_{k=0}^{k_\text{max}}\lambda_k(a,M_{\pi}) \nu^k+c_z(a,M_{\pi})z^2\right) \nonumber \\
{}\times{}&(1+c_a a^2+c_M(M_\pi^2-(M_\pi^{\text{phys}})^2),
\end{align}
We use $k_\text{max}=2$ here due to the level of noise. We present the results in Fig.~\ref{fig:RpITD_exta_1}. In the left panel of the figure, we show that the fit form reproduces the a09m310 and a12m220 data well. In the right panel we show the continuum-physical extrapolation expressed in Eq.~\ref{eq:RpITD-extra} with the solid band, along with the same fit form with the $a^2$ dependence replaced by $a$ for comparison in the dashed band. We see that our fit follows the data closely over the range of the plot. The $O(a)$ extrapolation has a similar mean value but larger error. Moving forward, we use the $a^2$ fit.

\subsection{Gluon PDF Results}
\begin{figure}[t!]
\centering
\includegraphics[width=0.48\textwidth]{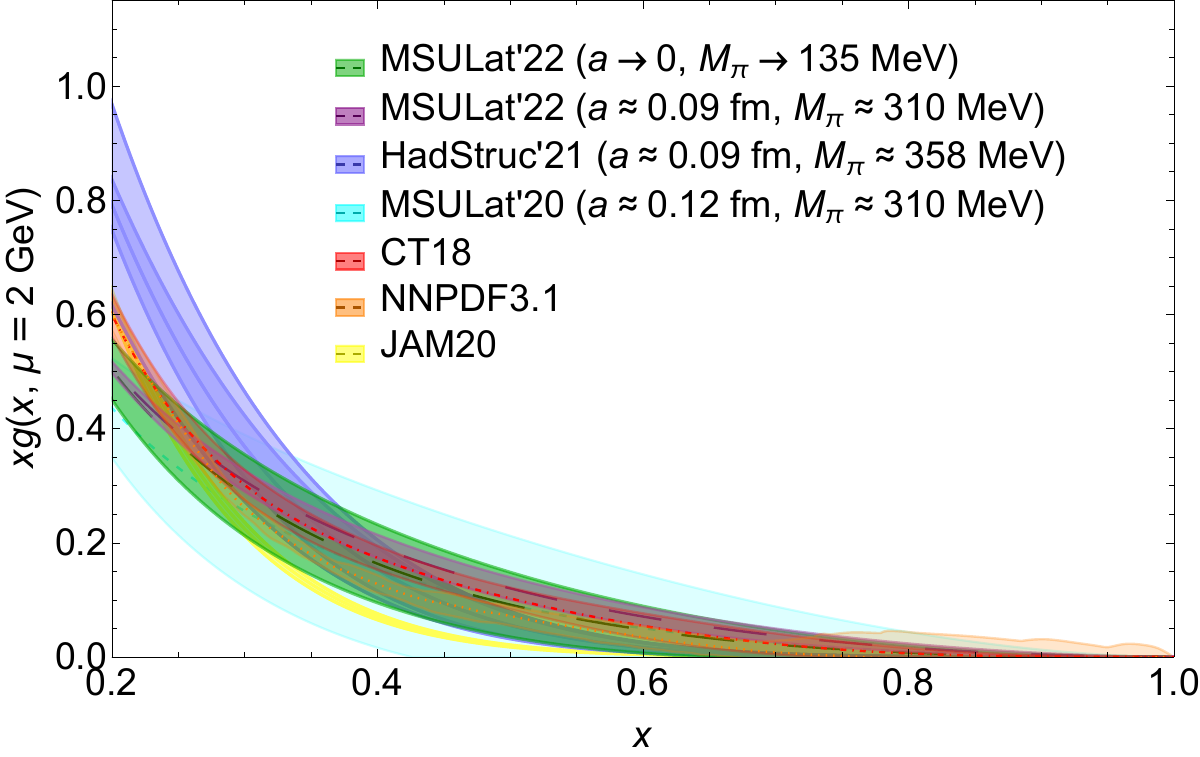}
\includegraphics[width=0.48\textwidth]{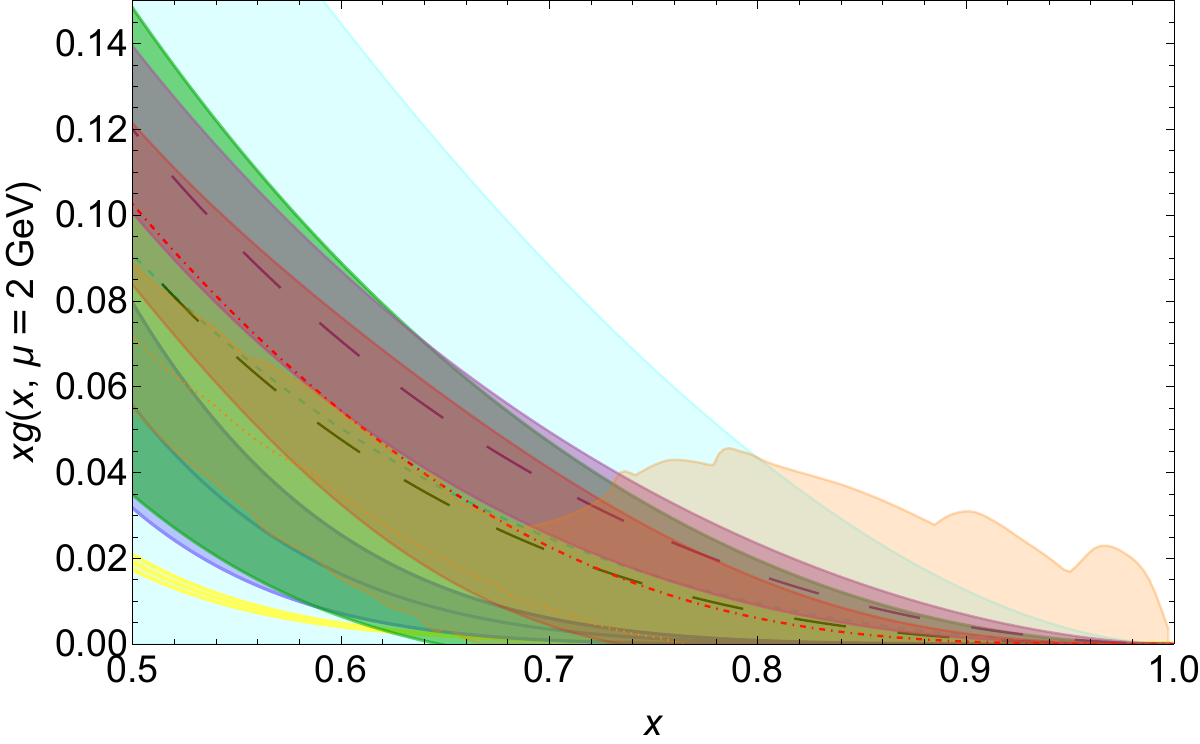}
\caption{
The unpolarized gluon PDF, $xg(x,\mu)$, with $x \in [0.2,1]$ (left) and $x \in [0.5,1]$ (right) obtained from our continuum-physical (green) and a09m310-ensemble (purple) RpITDs compared with the CT18 NNLO~\cite{Hou:2019efy} (red band),  NNPDF3.1 NNLO~\cite{Ball:2017nwa} (orange ban) and JAM20~\cite{Moffat:2021dji} (yellow band) gluon PDFs at $\mu=2$~GeV in the $\overline{\text{MS}}$ scheme.
Other prior lattice calculations of $xg(x)$ (including those done at single ensemble) from
HadStruc ($a\approx 0.094$~fm, $M_\pi \approx 358$~MeV)~\cite{HadStruc:2021wmh} (blue band) and MSULat~\cite{Fan:2020cpa} (cyan band) are also shown in the plot.
}
\label{fig:xg-comp}
\end{figure}

Using the physical-continuum extrapolated RpITD, we can extract the gluon PDF using the pseudo-PDF matching condition~\cite{Balitsky:2019krf}

\begin{equation}
\label{eq:matching-gg}
\mathscr{M}(\nu,z^2) = \int_0^1 dx \frac{xg(x,\mu^2)}{\langle x \rangle_g}R_{gg}(x\nu,z^2\mu^2),
\end{equation}
where $\mu$ is the renormalization scale in the $\overline{\text{MS}}$ scheme
and $\langle x \rangle_g=\int_0^1 dx \, x g(x,\mu^2)$ is the gluon momentum fraction of the nucleon. $R_{gg}$ is the gluon-in-gluon matching kernel described in Ref.~\cite{Balitsky:2019krf} and used in several other studies on the gluon PDF from lattice QCD~\cite{Fan:2020cpa,Fan:2021bcr,Salas-Chavira:2021wui,HadStruc:2021wmh}. We ignore the quark contributions, based on past study from our group showing that they are small~\cite{Fan:2021bcr}. 

We obtain $g(x,\mu^2)/\langle x \rangle_g$ by fitting the RpITD through the condition in Eq.~\ref{eq:matching-gg} using a fit form commonly used in global analyses:
\begin{equation}
f_g(x,\mu) = \frac{xg(x, \mu)}{\langle x \rangle_g(\mu)} = \frac{x^A(1-x)^C}{B(A+1,C+1)},
\label{functional}
\end{equation}
for $x\in[0,1]$ and zero elsewhere
The beta function $B(A+1,C+1)=\int_0^1 dx\, x^A(1-x)^C$ is used to normalize PDF properly.
Not shown here, we also considered a three-parameter fit, which produced results with a similar mean and just larger error bars. More on this is explored in the paper \cite{fan2022gluon}.

After fitting $f_g(x,\mu)$, the unpolarized nucleon gluon PDF $xg(x)$ can be extracted by taking the ratio of $f_g(x,\mu)=xg(x, \mu)/\langle x \rangle_g(\mu)$ and the gluon momentum fraction $\langle x \rangle_g(\mu)$ obtained in Ref.~\cite{Fan:2022qve}. We present our results in Fig.~\ref{fig:xg-comp} and compare to several other studies, including another lattice calculation \cite{HadStruc:2021wmh}. We see very good agreement with the other global fit and the Hadstruc21 \cite{HadStruc:2021wmh} PDFs in the large $x$-range.

\section{Summary and Conclusion}
We presented key information on the first continuum-limit unpolarized nucleon gluon PDF study described in further detail in Ref.~\cite{fan2022gluon}. We briefly describe the lattice setup with ensembles with three different lattice spacings and two different pion masses. We showed the extraction of the matrix elements from two point and three point correlators measured on the lattice. Then we described the extrapolation of the reduced Ioffe-time pseudo-distribution and presented the results of our light cone PDF matching compared to several other studies of the gluon PDF. 

We saw very good agreement of our PDF with those from other studies in the large $x$ region; however, the statistical errors in our results are still quite large. Future work should go towards smaller lattice spacings, larger boost momentum, and much higher statistics to improve our results in the small $x$ region. Further understanding of systematics may help obtain more precise results as well. 

\section*{Acknowledgments}

We thank MILC Collaboration for sharing the lattices used to perform this study. The LQCD calculations were performed using the Chroma software suite~\cite{Edwards:2004sx}.
This research used resources of the National Energy Research Scientific Computing Center, a DOE Office of Science User Facility supported by the Office of Science of the U.S. Department of Energy under Contract No. DE-AC02-05CH11231 through ERCAP;
facilities of the USQCD Collaboration are funded by the Office of Science of the U.S. Department of Energy,
and the Extreme Science and Engineering Discovery Environment (XSEDE), which was supported by National Science Foundation Grant No. PHY-1548562.
The work of  ZF and HL is partially supported by the US National Science Foundation under grant PHY 1653405 ``CAREER: Constraining Parton Distribution Functions for New-Physics Searches'' and by the  Research  Corporation  for  Science  Advancement through the Cottrell Scholar Award.  The work of WG is supported by MSU University Distinguished Fellowship.
The work of HL is partially supported by the US National Science Foundation under grant PHY 2209424.

\bibliography{refs.bib}
\end{document}